\documentclass[12pt]{iopart}
\usepackage{iopams}  
\usepackage{graphicx}
\begin{document}

\title{Viscous Hydrodynamics}

\author{Azwinndini Muronga$^{1,2}$}

\address{$^1$Gesellschaft f\"{u}r Schwerionenforschung (GSI). Planckstrasse 1,
D-64291 Darmstadt, Germany.}

\address{$^2$Institut f\"ur Theoretische Physik,
J.W. Goethe-Universit\"at,\\
Robert--Mayer--Stra{\ss}e 8--10, D--60325 Frankfurt am Main, Germany.}

\begin{abstract}
We study the role of viscosity in the early stages of relativistic
heavy ion collisions. We investigate the effects of viscosity on the
chemical equilibration of a parton gas. In the presence of viscosity the
lifetime of the system is increased. The temperature as well as the parton
fugacities evolves more slowly compared to ideal fluid dynamics.
\end{abstract}




\section{Introduction}
Ultra-relativistic nucleus-nucleus collisions probe the properties of nuclear
matter under extreme conditions \cite{QM}. Lattice quantum chromodynamics (QCD) 
 calculations \cite{Lattice} predict that ordinary
nuclear matter undergoes a phase transition to quark gluon plasma (QGP). An
important question is whether the high-energy density matter formed in
ultra-relativistic nuclear collisions lives sufficiently long enough to reach
thermodynamical equilibrium. That is does the matter reaches thermal, mechanical
and chemical equilibrium? In this work we assume that the matter reaches thermal
and mechanical equilibrium after proper time $\tau_0$.
We do not, however, assume that the matter is in chemical equilibrium.
Under these assumptions, and given initial values for temperature and the
quark, antiquark and gluon number densities, we can then employ fluid dynamics
to study the subsequent evolution of the kinetically equilibrated quark-gluon
phase, coupled to the rate equations which determine the chemical composition of
the system far away from equilibrium. This problem has been studied previously
in \cite{TSB,SMM,ER} using ideal fluid, in \cite{AKC} using first order
theory of dissipative fluid dynamics, and recently in \cite{XG} using parton
cascade.

Ideal (Euler) fluid dynamics has been useful in describing most of the observables at
RHIC \cite{HC,TH}.
In the early stages of heavy ion collisions, non--equilibrium effects play a 
dominant role. A complete description of the dynamics of heavy ion reactions
needs to include the effects of dissipation through 
non--equilibrium/dissipative fluid dynamics. As is well--known \cite{AM1,Aziz}, 
second order (or extended) theories (which are hyperbolic and causal) of dissipative fluids due 
to Grad \cite{HG}, M\"uller
\cite{IM}, and Israel and Stewart \cite{IS} were introduced
to remedy some undesirable features such as acausality. 
It seems appropriate
therefore to resort to hyperbolic theories instead of first order theories
(which are parabolic) in
describing the dynamics of heavy ion collisions. First order theories are 
due to Eckart
\cite{CE} and to Landau and Lifshitz \cite{LL} and they 
lead to Navier-Stokes-Fourier (NSF) equations which may be acausal. In addition
the first order theories do not have well-posed initial value problem.

\section{Viscous hydrodynamics and chemical equilibration}

In this work we extend the work of \cite{AM1} to include chemically
non-equilibrium effects. 
For partially equilibrated plasma of massless particles the equation of state
can be written as \cite{TSB}
\begin{equation}
\varepsilon = 3 p=\left[a_2\lambda_g+ b_2(\lambda_q+\lambda_{\bar{q}})\right]
T^{4} \enspace,
\end{equation}
where $a_2=8\pi^2/15$, $b_2=7\pi^2 N_f/40$ with $N_f$ being the number of quark
flavours, and the $\lambda_i$ are the parton fugacities defined through
\begin{equation}
n_g = \lambda_g n^{eq}_g,~~~~~~~~~~n_q = \lambda_q n^{eq}_q,
~~~~~~~~~~n_{\bar{q}} = \lambda_{\bar{q}} n^{eq}_{\bar{q}} \enspace,
\end{equation}
where the $n^{eq}_i$ are the equilibrium parton densities
\begin{equation}
n^{eq}_g = a_1 T^3~~~~~~~~~~n^{eq}_q = b_1 T^3 = n^{eq}_{\bar{q}}\enspace,
\end{equation}
where $a_1=(16/\pi^2)\zeta(3))$ and $b_1=(9/ 2\pi^2)\zeta(3)$.
The shear viscosity
coefficient is given by
\begin{equation}
\eta = \lambda_g \eta_g + \lambda_q \eta_q \enspace,
\end{equation}
where the shear viscosity coefficients for the quarks and gluons are given by 
\cite{HT,PA, HH} 
\begin{equation}
\eta_q = b_q T^3,~~~~~~~~~\eta_g = b_q T^3 \enspace,
\end{equation}
where $b_q = 0.82 (\alpha_s^2\ln(1/\alpha_s))^{-1}$ and 
$b_g = 0.20 ( \alpha_s^2\ln(1/\alpha_s))^{-1}$ 
with $\alpha_s$ being the strong coupling constant. 
We take $\alpha_s = 0.4$ throughout this analysis, unless otherwise stated.
 
In the absence of chemical equilibrium we need the master equations for the
evolution of parton densities. We consider only the dominant reactions
$gg\leftrightarrow ggg$, and $gg\leftrightarrow q\bar{q}$. 
For longitudinal boost invariant longitudinal flow under the assumed 
equation of state and transport coefficients the energy
equation and shear pressure \cite{AM1,AM3} become
\begin{eqnarray}
{\dot{\lambda}_g + b(\dot{\lambda}_q+\dot{\lambda}_{\bar{q}})\over
\lambda_g + b(\lambda_q+\bar{q})} +4{\dot{T}\over T} +{4\over 3 \tau} 
&=&  {1\over
\left[a_2\lambda_g+b_2(\lambda_q+\lambda_{\bar{q}})\right] T^4}{\Phi\over \tau} \enspace,\\
{{\rm d} \Phi\over {\rm d}\tau}
+{2\over 9}{\left[a_2\lambda_g+b_2(\lambda_q+\lambda_{\bar{q}})\right]
\over \left[\lambda_g b_g+\lambda_q b_q\right]} T\Phi
 &=& -{1\over 2}\Phi\left[{1\over\tau}-\left[5{\dot{T}\over T} +{\dot{\lambda}_g + b(\dot{\lambda}_q+\dot{\lambda}_{\bar{q}})\over
\lambda_g + b(\lambda_q+\bar{q})}\right]\right]  \nonumber\\
&&~~~~~
+ {8\over 27}{\left[a_2\lambda_g+b_2(\lambda_q+\lambda_{\bar{q}})\right]T^4\over
\tau} \enspace,
\end{eqnarray}
and are coupled to the master equations for the fugacities \cite{TSB,SMM,ER}
\begin{eqnarray}
{\dot{\lambda}_g\over \lambda_g}  +3{\dot{T}\over T} + {1\over \tau} 
&=& R_3(1-\lambda_g) - 2
R_2\left(1-{\lambda_q\lambda_{\bar{q}}\over \lambda_g^2}\right) \enspace,\\
{\dot{\lambda}_q\over \lambda_q}  +3{\dot{T}\over T} + {1\over \tau} 
&=& R_2 {a_1\over b1}\left({\lambda_g\over\lambda_q}-{\lambda_{\bar{q}}\over
\lambda_g}\right) \enspace,
\end{eqnarray}
where $b=b_2/a_2 =21 N_f/64$ with $N_f$ being the number of quark flavours
The reaction rates $R_2$ and $R_3$ are given by \cite{TSB}
\begin{equation} \label{Rs}
{\rm R}_2 \simeq 0.24\, N_f\,\alpha_s^2\, \lambda_g \,
T \, \ln \left( 1.65/\alpha_s\, \lambda_g \right) \,\,, 
\,\,\,\, {\rm R}_3 \simeq 2.1 \,\alpha_s^2 \, T \, \sqrt{2\lambda_g - 
\lambda_g^2} \ \ .
\end{equation}
We consider initial conditions relevant for RHIC: $\tau_0=0.25$ fm/c, 
$T_0=0.66$ GeV, $\Phi_0=p_0/5$, $\lambda_{g0}=0.34$, and $\lambda_{q0}=0.064$;
and for LHC: $\tau_0=0.25$ fm/c, 
$T_0=1.0$ GeV, $\Phi_0=p_0/5$, $\lambda_{g0}=0.43$, and $\lambda_{q0}=0.082$.
Here $p_0$ is the initial pressure. These sets of initial conditions are
motivated in \cite{SMM,AM1}.

\section{Results}
\begin{figure}[htb]
\vspace{0.3255cm}
\centerline{
     \includegraphics[width=7cm,height=4.5cm]{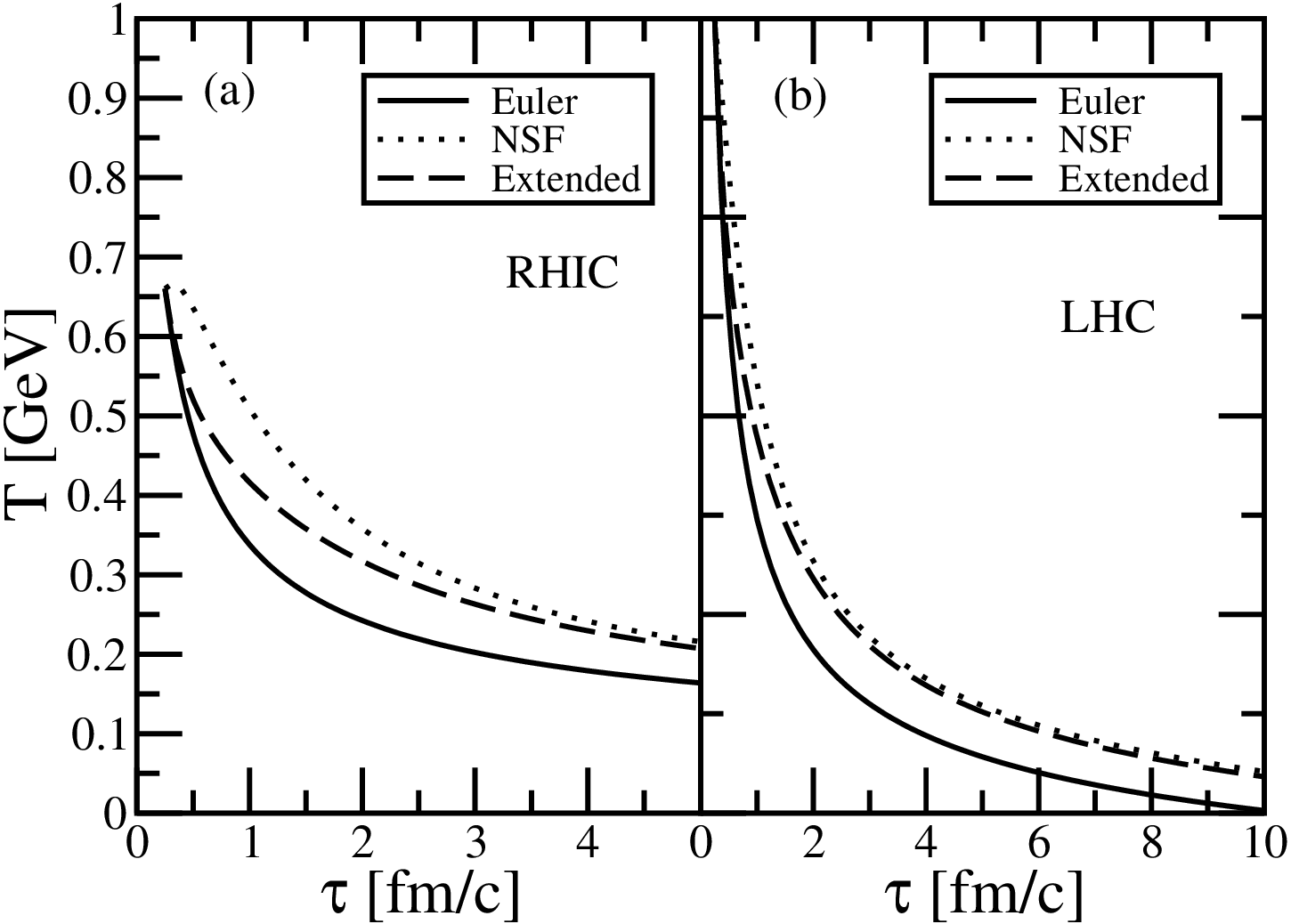}}
      \caption{Time evolution of temperature for (a) RHIC and (b) 
               LHC initial conditions. The curves are for  ideal fluid (solid), dotted
	       first order theory (dotted) and second order theory (dashed).}
        \label{fig:temp}
\end{figure}

In Fig. \ref{fig:temp}(a,b) we show the well known results (see \cite{AM1})
of the temperature evolution. The ideal fluid dynamics approximation leads to
faster cooling. Due to the reduction of longitudinal pressure, less work is done
to the expansion and hence the slow cooling in the presence of viscosity.
However the first order theory even predicts heating during the expansion stage.
This is in contradiction to the energy conservation laws. Also the first order
theory will overestimate the freeze-out temperatures. This in turn might lead to
wrong conclusions about the observables. On the other hand the second order
theory does not have these undesirable features that are exhibited by first
order theory.

\begin{figure}[htb]
\vspace{0.3255cm}
\begin{minipage}[t]{80mm}
\centerline{
     \includegraphics[width=7cm,height=4.5cm]{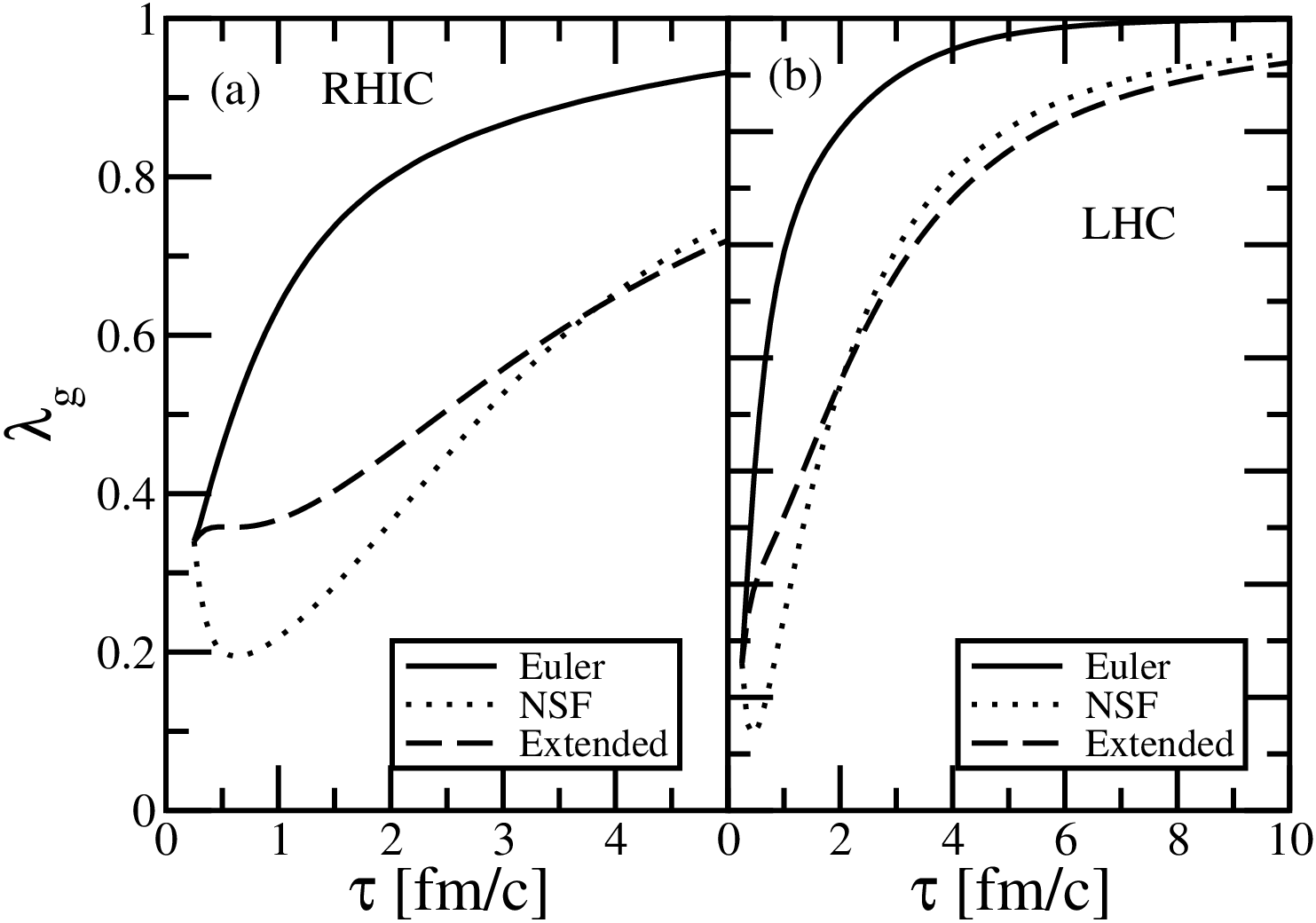}}
      \caption{Time evolution of gluon fugacity for (a) RHIC and (b) 
               LHC initial conditions.}
        \label{fig:gfug}

\end{minipage}
\vspace{0.2 cm}
\begin{minipage}[t]{80mm}
\centerline{
     \includegraphics[width=7cm,height=4.5cm]{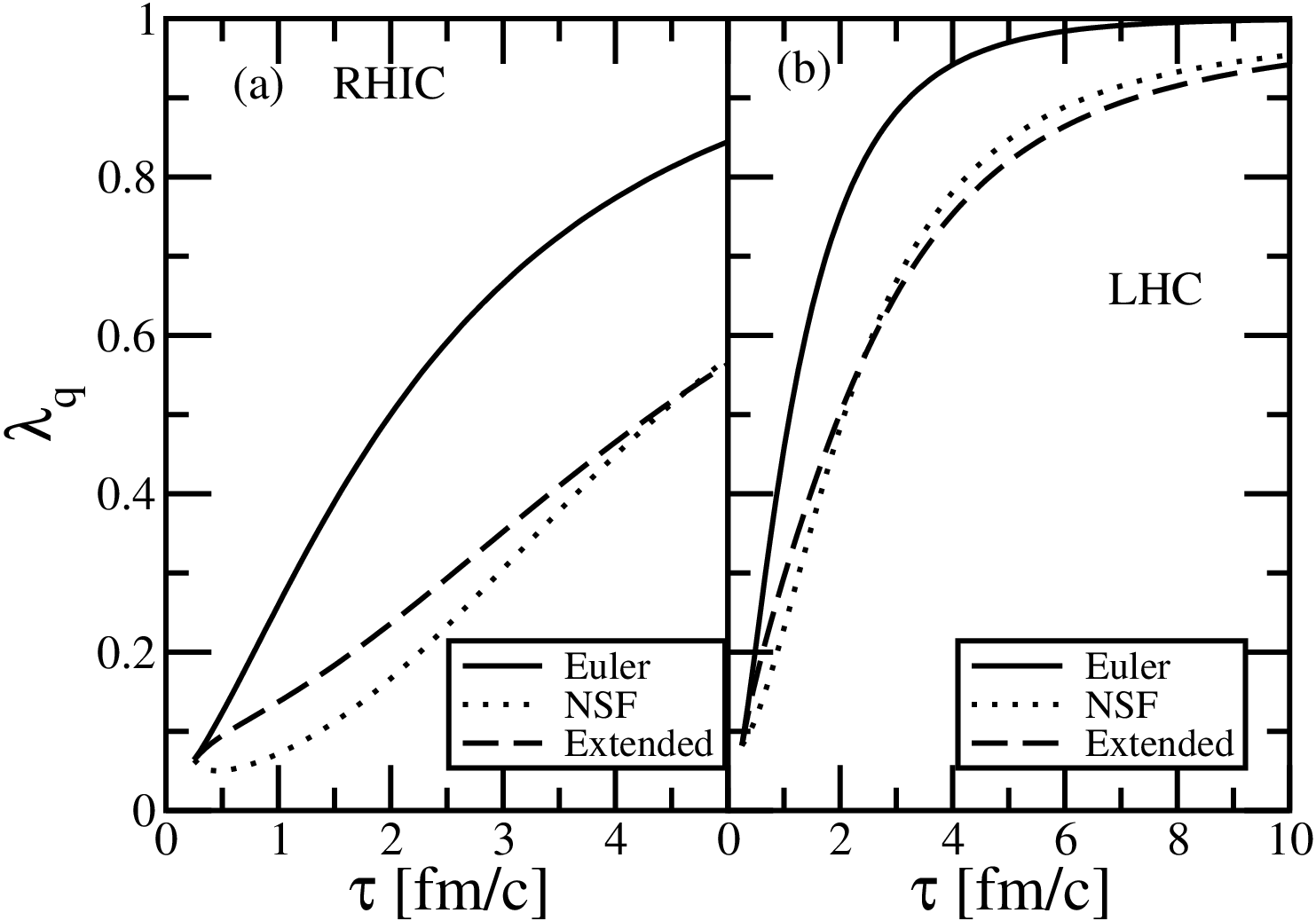}}
      \caption{Time evolution of quark fugacity for (a) RHIC and (b) 
               LHC initial conditions.}
        \label{fig:qfug}
\end{minipage}

\end{figure}

In Figs. \ref{fig:gfug}(a,b) and \ref{fig:qfug}(a,b)  we show time evolution of the parton
fugacities. In the presence of viscosity the parton viscosities evolve more
slowly towards chemical equilibration. This will have considerable effects on
the observables such as strangeness production. In the early stages of the
expansion of the system one sees the difference between first order theory and
second order theory. This difference is investigated in more detail \cite{AM3}.                                                                                                                                                                                                              

\section{Conclusions}
We have investigated the effects of shear viscosity on the chemical
equilibration of the parton system in relativistic nuclear collisions. Due to
slow cooling of the system in the presence of viscosity chemical equilibration
is slowed.
The effects of transverse expansion \cite{AM2} and of the mass of strange quark 
in the chemically
non-equilibrated viscous system are being studied and will be published somewhere
\cite{AM3}.


\begin{thebibliography}{9}
\bibitem{QM}
	A compilation of current RHIC results can be found in: 
	Quark Matter '04.
\end{thebibliography}
\end{document}